\documentclass[10pt,letterpaper]{article}
\usepackage{opex3}

\begin{document}

\title{Generation of paired photons in a quantum separable state in Bragg reflection waveguides}

\author{Ji\v{r}\'{i} Svozil\'{i}k$^{1,2}$, Martin Hendrych$^{1}$,  Amr S.
Helmy$^{3}$,
and Juan P.
Torres$^{1,4}$ }

\address{$^1$ ICFO-Institut de Ciencies Fotoniques, Mediterranean
Technology Park, 08860 Castelldefels (Barcelona), Spain}

\address{$^2$ Joint Laboratory of Optics, Palack\'{y} University
and Institute of Physics of Academy of Science of the Czech
Republic, 17. listopadu 50A, 772 07 Olomouc, Czech Republic}

\address{$^3$ Edward S. Rodgers Department of Electrical and Computer
 Engineering, University of Toronto, 10 King's Collage Road, Toronto, Ontario 
 M5S3G4, Canada}
\address{$^4$ Departament of Signal Theory and Communications, Universitat
Politecnica de Catalunya, 08034 Barcelona, Spain}

\begin{abstract}
This work proposes and analyses a novel approach for the generation of separable
(quantum uncorrelated) photon pairs based on spontaneous
parametric down-conversion in Bragg reflection waveguides composed
of semiconductor AlGaN layers. This platform allows the removal of any spectral
correlation between
paired photons that propagate in different spatial modes. The
photons can be designed to show equal or different spectra
by tuning the structural parameters and hence the dispersion of the waveguide.
\end{abstract}


\email{jiri.svozilik@icfo.es}

\section{Introduction}
Spontaneous parametric down-conversion (SPDC) in $\chi^{(2)}$
nonlinear media is a well-known parametric process that allows the
generation of photon pairs. In this process, the interaction of an
intense pump beam with the atoms of a nonlinear medium mediates
the generation of pairs of photons with lower frequency. The
spatio-temporal properties of the down-converted photons depend on
the specific SPDC configuration considered as well as on the
spatial and temporal characteristics of the pump beam.

In most applications the goal of using SPDC is the generation of
entangled photon pairs. However, the generation of photon pairs that lack
any entanglement (quantum separability), but are generated in the
same time window, is also of paramount importance for quantum
networking and quantum information processing
\cite{rohde2005,walmsley2005,kok2007}. By and large, separable
photon pairs are not harvested directly at the output of the
down-converting crystal \cite{clara:2010}. Their generation
in a separable quantum state requires intricate control
of the properties of the down-converted photons in all the degrees
of freedom; namely polarization, spatial profile and frequency content. In
most cases, the separability in polarization results directly
from the phase matching conditions imposed by the specific SPDC
configuration chosen. Separability in the spatial degree of
freedom can readily be imposed by an appropriate projection of the
down-converted photons into specific modes. Using
single-mode fibers is one example. This leaves separability in the frequency
degree of freedom as a significant challenge to achieve quantum
separability between the down-converted photons.

Although one can always resort to strong spectral filtering to enhance the
quantum separability of the two-photon state \cite{aichele2002},
this entails a substantial reduction in the brightness of the
photon source. In the past few years, several methods have been
proposed and implemented to generate frequency-uncorrelated 
photon pairs without the need for strong filtering. For example,
elimination of the frequency correlation of photon pairs can be
achieved when the operating wavelength, the nonlinear material and its length
are  appropriately chosen \cite{grice2001}, as  has
been demonstrated in \cite{mosley2008}.

It is important to note that the appropriate conditions for the generation of
frequency-uncorrelated photons might not always exist or might not
be convenient for a given application \cite{mosley2008}. The use of
achromatic phase
matching, or tilted-pulse techniques, allows the generation of
separable two-photon states independently of the specific
properties of the nonlinear medium and the wavelength used. This
method employs the Poynting vector walk-off exhibited by nonlinear
crystals outside noncritical phase matching to modify the
effective group velocity of all the interacting waves, therefore
allowing for the control of the frequency correlation
\cite{torres2005,hendrych2007}. This is a route by which frequency-uncorrelated
photons can be generated \cite{torres2010}.

Non-collinear SPDC also allows the control of the generation of
frequency-uncorrelated photons by controlling the pump-beam width
and the angle of emission of the down-converted
photons~\cite{walton2003,uren2003}. It is indeed possible to map
the spatial characteristics of the pump beam into the spectra of
the generated photons (spatial-to-spectral mapping)
\cite{carrasco2004}, thus providing another way to manipulate the joint
spectral amplitude of the biphoton, as has been demonstrated in
\cite{valencia2007}. The combination of using the pulse-tilt techniques
described above together with using non-collinear geometries further
expands the possibilities to control the joint spectrum of photon pairs
\cite{shi2008}.
Another approach to control the frequency correlations is to use nonlinear
crystal 
superlattices \cite{uren2006}.

The method presented here to generate frequency-uncorrelated
photons is based on tuning the dispersive properties of the
nonlinear medium itself by engineering the waveguide dispersion. The waveguide
geometry allows us to enhance the source brightness and
to control the dispersive properties of the propagating mode by tailoring the
contribution of the waveguide dispersion to the overall
dispersion. The use of Bragg reflection waveguides (BRW) based on
III-V ternary semiconductor alloys (Al$_x$Ga$_{1-x}As$,
Al$_x$Ga$_{1-x}N$) offers several advantages over conventional
materials such as a large nonlinear coefficient, a broad transparency
window and mature semiconductor fabrication technologies that can be utilized to
tailor the spatio-temporal properties of the generated photon pairs.

Although phase-matching (PM) in these materials has often been
considered as challenging due to their lack of birefringence as
well as their highly dispersive nature, one can take advantage of
the different dispersive properties of the various modes the
waveguide can support; namely totally-internally-reflected (TIR) modes and Bragg
modes. In this case,
one can fully exercise a significant level of control over the
dispersion properties of the interacting waves while operating at
the phase--matching condition \cite{west2006}. This property has
been successfully used to enhance the capabilities of BRWs for
generating frequency-anticorrelated photon pairs with a tunable
bandwidth \cite{abolghasem2009}. To even further expand the
freedom of design of the dispersive properties, and thus broaden
the set of possible properties of the generated photons, 
quasi-phase-matching, with the periodic reversal of the sign
of the nonlinear coefficient $\chi^{(2)}$, can also be used in conjunction.

In this paper we demonstrate that Bragg reflection waveguides made of
Al$_x$ Ga$_{1-x}$ N slabs can be tailored to generate photon pairs in a quantum
separable state. Quasi-phase-matching (QPM) of the
core slab is used to satisfy the phase-matching condition, while the tailoring
of the dispersive properties of the waveguide, enhanced by using different types
of modes for each of the interacting waves, allows us to control the
frequency correlations between the down-converted photons.

\section{Description of the quantum state of the down-converted photons}
The quantum state of the down-converted photons (the signal
and idler) at the output face of the waveguide, while neglecting the vacuum
contribution, can be written as
\begin{equation}
|\Psi\rangle=\int d\Omega_{s}
d\Omega_{i}\Phi(\Omega_{s},\Omega_{i})
\hat{a}_{s}^{\dagger}(\omega_s^0+\Omega_{s})\hat{a}_{i}^{\dagger}
(\omega_i^0+\Omega_{i})
|0\rangle_s |0\rangle_i,
\end{equation}
where $\hat{a}_{s}^{\dagger}(\omega_s+\Omega_{s})$ and
$\hat{a}_{i}^{\dagger}(\omega_i+\Omega_{i})$ designate the
creation operators of signal and idler photons at frequencies
$\omega_s^0+\Omega_s$ and $\omega_i^0+\Omega_i$, respectively.
$\omega_s^0=\omega_i^0$ are the central frequencies of the signal
and idler photons, and $\Omega_{s,i}$ designate the frequency
deviations from the corresponding central frequencies. The signal
and idler photons are generated in specific spatial modes of the
waveguide as will be described later.

The biphoton amplitude $\Phi(\omega_{s},\omega_{i})$ is given by
\begin{equation}
\Phi(\Omega_{s},\Omega_{i})={\cal N} E_{p}(\omega_{s}+\omega_{i})
\textrm{sinc}\left(\frac{\Delta_k L}{2} \right)\exp \left( i
\frac{s_k L}{2} \right),
\end{equation}
where $\Delta_k=k_{p}-k_{s}-k_{i}$ and $s_k=k_p+k_s+k_i$.
$k_{p,s,i}$ are the longitudinal ($z$) components of the wavevector of all the
interacting photons. $E_p$ is the spectral
amplitude of the pump beam of central frequency $\omega_p^0
=\omega_s^0 + \omega_i^0 $ at the input face of the waveguide,
which is assumed to be Gaussian. As such,
$E_{p}(\Omega_{p})\sim\exp\left(-\Omega_{p}^2/\Delta\omega_{p}^{2}\right)$.
${\cal N}$ is a normalizing constant, which ensures that $\int\int
d\Omega_{s}d\Omega_{i}
|\Phi(\Omega_{s},\Omega_{i})|^2=1$.

\begin{figure}[t]
  \centering
   \includegraphics[scale=0.9]{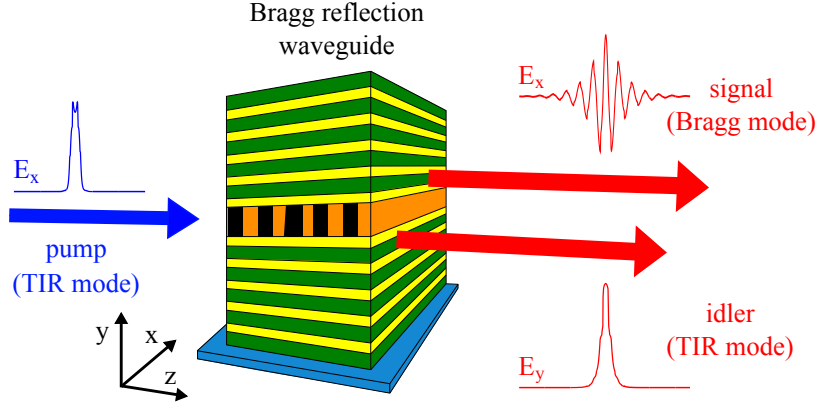}
  \caption{General scheme for the generation of frequency-uncorrelated
  photon pairs. The Bragg and TIR modes have different group velocities which
can be
  properly engineered by modifying the waveguide structure. }
  \label{fig1}
\end{figure}

In BRWs, photons can be guided by the conventional total internal
reflection (TIR modes) or by a distributed reflection from the
transverse Bragg reflectors (Bragg modes). The spatial modes of
the pump, signal and idler photons that we will consider here are
shown schematically in Fig.~\ref{fig1}. The pump and idler photons
propagate as a TIR mode. The signal photons propagate as a Bragg
mode. The use of different spatial modes for the signal and idler
enhances the control of the dispersive properties of the SPDC
process.

In order to get a further insight into the procedure to search for
BRW configurations that generate separable paired photons, we make
use of the fact that the dispersive properties of the interacting
waves are different and expand their longitudinal wavevectors to
first order, so that $k_{j}=k_{j}^{0}+N_{j} \Omega_{j}$ with
$j=p,s,i$. $k_j^{(0)}$ are the longitudinal wavevectors at the
central frequencies $\omega_j^0$, and $N_j$ are the inverse group
velocities. Under these conditions, the biphoton amplitude can be
written as
\begin{eqnarray}
\label{uncorrelated} & & \Phi(\Omega_{s},\Omega_{i})={\cal N}
\exp\left\{ -\frac{\left(\Omega_{s}+\Omega_{i}\right)^2}{\Delta
w_p^2} \right\} \textrm{sinc}\left\{ \left[\left(N_p-N_s\right)
\Omega_s+\left(N_p-N_i\right) \Omega_i \right]\frac{L}{2}\right\}
\nonumber \\
& &  \times \exp \left\{ i \left[(N_p+N_s)\Omega_s+ (N_p+N_i)
\Omega_i \right] \frac{L}{2}\right\}.
\end{eqnarray}
Upon inspecting Eq. (\ref{uncorrelated}), one can  show that if the inverse
group velocities of the signal (idler) and pump are equal $N_p=N_s$ ($N_p=N_i$),
then increasing the bandwidth of the pump beam bandwidth such that
$\Delta \omega_p \gg 1/|N_p-N_{s,i}|L$ allows us to erase all the
frequency correlations between the signal and idler photons.
Notice that in this case, even though there is no entanglement
between the signal and idler photons, the bandwidth of one the photons
is larger than the bandwidth of the other photon. The
quantum state is separable but the photons are distinguishable by
their spectra.

To generate uncorrelated and indistinguishable photon pairs, the
condition $N_p=(N_s+N_i)/2$ should be fulfilled together with the
condition for the bandwidth
\begin{equation}
\Delta \omega_p \simeq \frac{2}{\alpha L
 \sqrt{N_s-N_p}\sqrt{N_p-N_i}}.
\end{equation}
This condition is obtained from approximating the sine cardinal function
$\textrm{sinc(x)}$
 in Eq. (\ref{uncorrelated}) by a Gaussian function $\exp{\mathrm{[-(\alpha
x)^2]}}$ with
$\mathrm{\alpha=0.439}$.

To quantify the degree of entanglement of the generated two-photon
state, we calculate the Schmidt decomposition of the biphoton
amplitude, i.e., $\Phi(\Omega_s,\Omega_i)=\sum_{n=0}^{\infty}
\sqrt{\lambda_{n}} U_n(\Omega_s) V_n(\Omega_i)$, where $\lambda_n$
are the Schmidt eigenvalues and $U_n$ and $V_n$ are the
corresponding Schmidt modes. The degree of entanglement of the
two-photon state can be quantified by means of the purity of
either of the subsystems (signal or idler photons) that make up the whole
system.  The purity of either subsystem is given
by $P=K^{-1}$, where $K=\sum_{n=0}^{\infty} \lambda_n$. $K=1$
corresponds to a separable two-photon state, while an increasing
value of $K$ corresponds to an increase in the degree
of entanglement.

\begin{figure}[b]
\begin{center}

\begin{tabular}{cccc}

\begin{minipage}[c]{0.25\textwidth}%
\begin{center}
\includegraphics[scale=0.38]{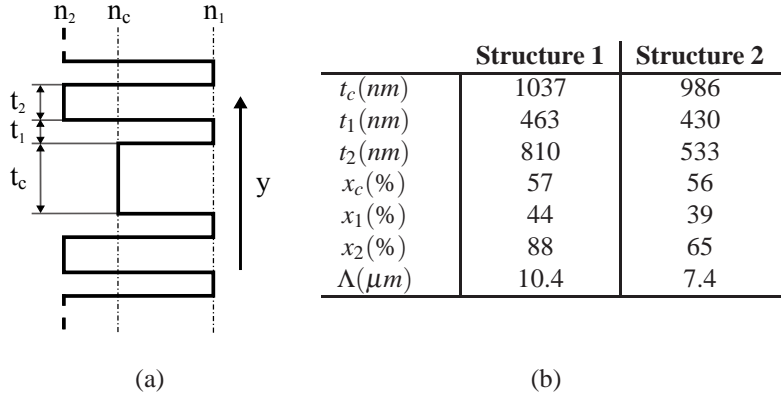}
\par\end{center}%
\end{minipage} \hspace{0.3cm}

&
\begin{minipage}[c]{0.45\textwidth}%
\begin{center}
\begin{tabular}{cc|c|c}
&
\multicolumn{1}{c}{} & \multicolumn{1}{c|}{\textbf{Structure 1}} &
\textbf{Structure 2}\tabularnewline
\hline
$t_{c}$$(nm)$  &  & 1037  & 986\tabularnewline
$t_{1}$$(nm)$  &  & 463  & 430\tabularnewline
$t_{2}$$(nm)$  &  & 810  & 533\tabularnewline
$x_{c}(\%)$  &  & 57  & 56\tabularnewline
$x_{1}(\%)$  &  & 44  & 39\tabularnewline
$x_{2}(\%)$  &  & 88  & 65\tabularnewline
$\Lambda$$(\mu m)$ & & 10.4 & 7.4 \tabularnewline
\hline
\end{tabular}
\par\end{center}%
\end{minipage}\tabularnewline
\tabularnewline
(a) & (b)
\end{tabular}
\end{center}

\caption{(a) Profile of the refractive index along the
\emph{y}-axis of the Bragg reflection waveguide. $t_{c}$ - core
thickness; $t_{1,2}$ - thicknesses of the alternating layers of
the Bragg reflector; $x_{c}$ - aluminium concentration in the
core; $x_{1,2}$ - aluminium concentration in the reflector's
layers; $\Lambda$ - quasi-phase-matching period. Both structures are 4 mm long
and they are optimized for type-II SPDC.}

\end{figure}

\section{Design of BRW structures to generate uncorrelated photon pairs}
Let us consider the generation of paired photons in the C-band of the optical
communication window, i.e., let the central wavelength of both
emitted photons be 1550 nm. Therefore, for the
frequency-degenerate case, the central wavelength of the pump beam
is 775 nm. The main parameters that characterize the dispersion
properties of the Bragg modes, and that can be engineered to
tailor the spectral properties of the down-converted photons, are the
thickness of the layers and their aluminium fraction.

BRW structures for the generation of frequency-uncorrelated photon
pairs were obtained by numerically solving the Maxwell equations
inside the waveguide using the finite element method
\cite{Jianming2002FEM}. Since many solutions were found, a genetic
algorithm was used to select waveguides with the properties that are most
suitable for practical implementation. The thicknesses and the corresponding
aluminum fractions of two of the structures obtained are given  
in Fig. 2.

%

The refractive indices for the calculations were taken
from \cite{laws:1108}. The Bragg reflection waveguides are composed
of 12 bi-layers above and below the core. Both structures were
optimized for the Bragg mode propagation at the quarter-wave
condition for the central wavelength, which maximizes the energy
confinement in the core.

Type-II SPDC interactions are considered for both structures, even though
structures with type-I or type-0 interactions can also be designed. One of the
advantages
of type-II phase-matching is that the generated photons can easily be
separated at the output of the waveguide. The pump and signal photons have TE
polarization and the idler photons have TM polarization. The signal photons
propagate as a Bragg 
mode, whereas the idler photons propagate as a TIR mode. The
quasi-phase-matching can be achieved, for example, by the method described in
\cite{pezzagna:062106}. The quasi-phase-matching periods $\Lambda$ were
calculated from the
phase-matching condition $\Delta_{k}-\frac{2\pi}{\Lambda}=0$, where $\Delta_{k}$
is the
phase-mismatch function at the central frequencies of all interacting waves.

The spatial overlap between the modes of the interacting photons is
defined as
\begin{equation}
\Gamma=\int dx\, u_p(x)u^*_s(x)u^*_i(x),
\end{equation}
where $u_j(x)$ $j=p,s,i$ are the mode functions describing the
transverse distribution of the electric field in the waveguide.
The overlap reaches $40.5\%$ for Structure
1 and $19.4\%$ for Structure 2. The combination of the high effective nonlinear
coefficient and the overlap results in an efficiency that is still much
higher than with other phase-matching platforms in waveguides or bulk
media.  Although the thickness of the core of both structures is sufficiently
large so that higher-order modes (both TIR and Bragg modes) could exist,
they lack phase-matching and their overlap is very small. 

\begin{figure}[b]
\begin{centering}
 \includegraphics[scale=0.8]{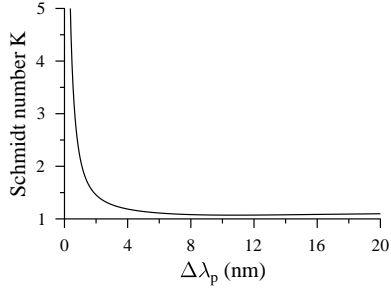}
\par\end{centering}
\caption{The Schmidt number K as a function of the bandwidth of
the pump beam $\Delta\lambda_{p}$.}
\end{figure}
\begin{figure}[t]
\centering
\begin{tabular}{cc}

\includegraphics[scale=0.65]{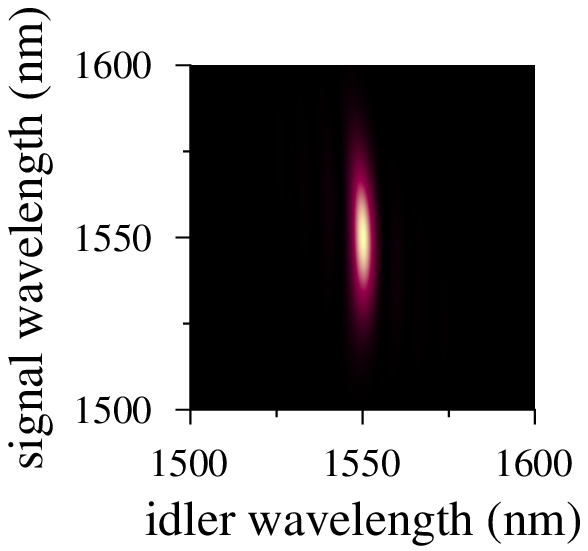}
&
\includegraphics[scale=0.65]{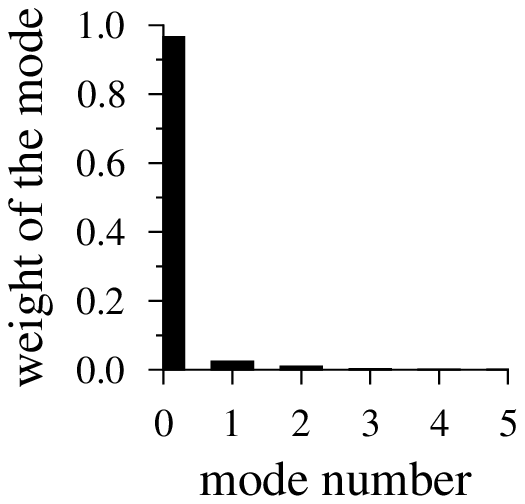} \tabularnewline
(a) &(b)
\end{tabular}
\caption{(a)
Joint spectral intensity of the biphoton generated in Structure 1
for $\Delta\lambda_{p}$=10 nm. (b) The Schmidt decomposition
corresponding to this quantum state.}
\end{figure}

\begin{figure}[b]
\centering{}\includegraphics[scale=0.8]{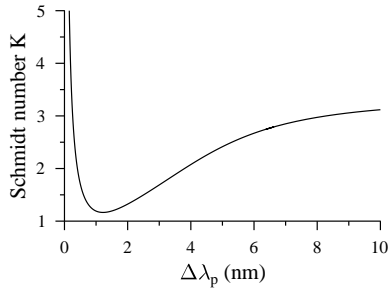}\caption{The Schmidt
number K of the state generated in Structure 2 as a function of
pump bandwidth $\Delta\lambda_{p}$.}

\end{figure}

\begin{figure}[t]
\centering
\begin{tabular}{ccc}
\includegraphics[scale=0.65]{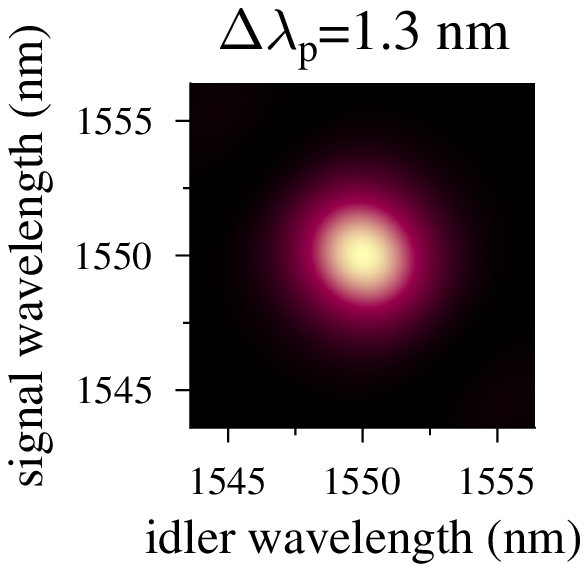}
&
\includegraphics[scale=0.65]{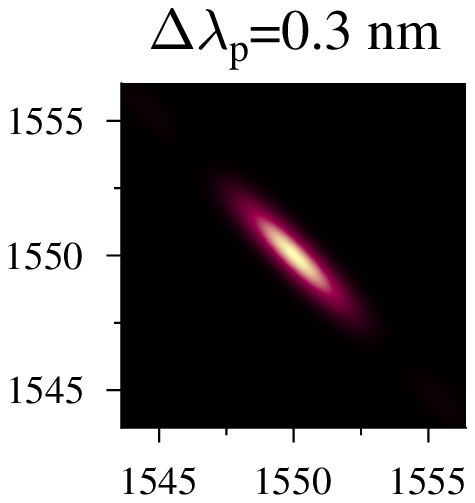}
&
\includegraphics[scale=0.65]{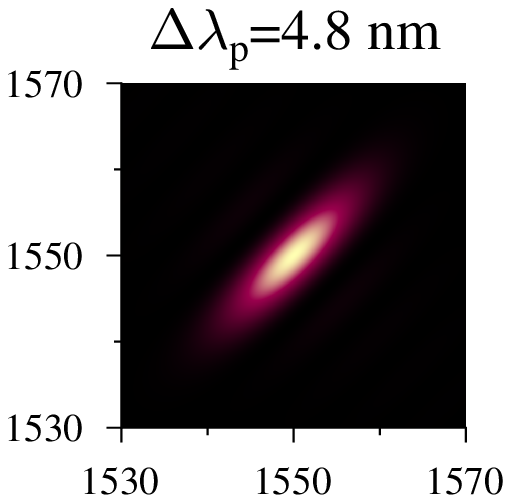}
\tabularnewline
(a) &(b) & (c)
\tabularnewline
\includegraphics[scale=0.65]{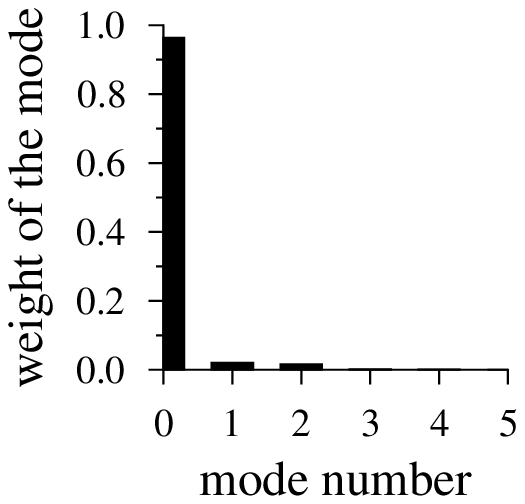}
&
\includegraphics[scale=0.65]{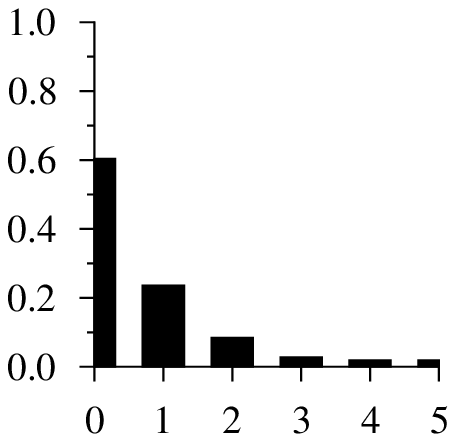}
&
\includegraphics[scale=0.65]{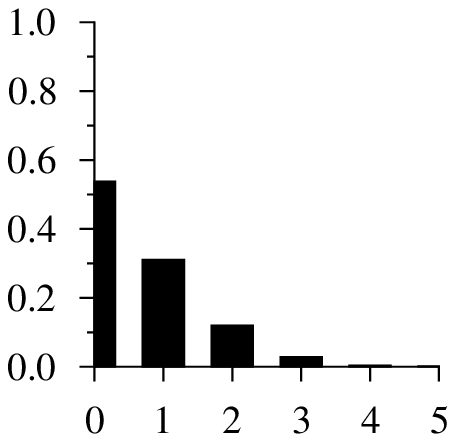}
\tabularnewline
(d) &(e) & (f)
\end{tabular}
\caption{Joint spectral intensity of photons generated in Structure 2 for
different pump bandwidths: a) $\Delta\lambda_{p}=1.3$ nm, b)
$\Delta\lambda_{p}=0.3$ nm and c) $\Delta\lambda_{p}=4.8$ nm.
Plots (d), (e) and (f) in the second line are the corresponding
Schmidt decompositions.}

\end{figure}

\subsection{Uncorrelated photon pairs with different spectra}
Structure 1 provides a configuration to generate a quantum
separable state  with different
spectral bandwidths of the signal and idler photons. The group
velocities of the pump and signal photons are equal. We find that
$\mathrm{v}_{p}=\mathrm{v}_s=0.445c$, where $c$ is the speed of light in
a vacuum.
The dependency of the Schmidt number K on the pump beam bandwidth is plotted in
Fig.~3. A highly separable quantum state can be obtained for a pump beam 
bandwidth $\Delta\lambda_{p}\ge 10$ nm. For values of
$\Delta\lambda_{p} < 1$ nm, the paired photons turn out to be
anti-correlated.

The joint spectral intensity of the biphoton is showed in
Fig.~4(a). It shows a cigar-like shape oriented along the signal
wavelength axis, as expected from the fulfillment of the condition
$N_p=N_s$. The Schmidt decomposition is shown in Fig.~4(b).
Clearly, this decomposition corresponds to a nearly ideal case of
frequency-uncorrelated photons. For the case shown in Fig.~4, with
a pump beam bandwidth (FWHM) of 10 nm, the bandwidths of
the signal and idler photons are 47.5 nm and 8 nm, respectively.
 The entropy of entanglement is used as a measure of
spectral correlation \cite{Nielsen} and is defined
as $E=-\sum_i\lambda_i\log_2\lambda_i$. The obtained
value is 0.257 in this case.

\subsection{Uncorrelated photon pairs with identical spectra}
Structure 2 is designed for the generation of a separable
two-photon state where both photons exhibit the same spectra. The 
calculated values of the group velocities of all the waves are $\mathrm{v}_{p} =
0.44 c$, $\mathrm{v}_{s} = 0.4252 c$ and
$\mathrm{v}_{i} = 0.456 c$. Figure~5 shows the value of the Schmidt number K as
a function of the pump beam
bandwidth. The optimum pump bandwidth for the generation of
frequency-uncorrelated photons is found to be
$\Delta\lambda_{p}=1.3$ nm, for which K achieves its lowest value. The value
of K cannot reach the ideal value of 1 due to the presence of the side-lobes of
the sinc
function in the anti-diagonal direction and a Gaussian profile in the diagonal
direction 
that introduces a slight asymmetry (see Eq.~3). Figure~6(a) shows the joint
spectral intensity of
frequency-uncorrelated photons, when this optimum value of the
pump bandwidth is used. Figure~6(d) shows the corresponding Schmidt
decomposition. The entropy of entanglement is 0.267 and the
bandwidth is 4.5 nm for both signal and idler photons.
  
For smaller values of the pump beam bandwidth, the photons
generated in Structure 2 correspond to photon pairs that are
anticorrelated in frequency (see Fig.~6(b)), whereas the use of
larger values allows the generation of frequency-correlated
photon pairs (see Fig.~6(c)). Figures 6(e) and (f) show the Schmidt
decompositions corresponding to each of these cases.

\section{Conclusion}
We have presented and analyzed a new source for photon pairs that allows the 
generation of paired photons that lack any frequency correlation. These are of
paramount importance for quantum networking technologies and
quantum information processing. The source is based on Bragg
reflection waveguides composed of Al$_x$Ga$_{1-x}N$.
Quasi-phase-matching of the waveguide core is used to achieve
phase-matching at the desired wavelength. The control of
BRW dispersion is used to control the frequency correlation
between the generated photons.

Two Bragg reflection waveguide structures have been presented. One of the
structures allows us to generate uncorrelated photons with different spectra.
The down-converted uncorrelated photons generated in the second structure are
spectrally indistinguishable.

This technique offers a promising route for the realization of
electrically pumped, monolithic photon--pair sources on a chip
with versatile characteristics.

\section{Acknowledgments}
This work was supported by the Government of Spain (Consolider
Ingenio CSD2006-00019, FIS2010-14831). This work was also
supported in part by FONCICYT project 94142 and by projects IAA100100713 of
GA AV\v{C}R, COST OC 09026.
\end{document}